\documentclass[twocolumn,floatfix,prl,showpacs,superscriptaddress]{revtex4}
\usepackage{graphicx}

\begin{document}
\title{Electron Spin Resonance in sine-Gordon spin chains in the
perturbative spinon regime}
\author{S. A. Zvyagin}
\affiliation{Hochfeld-Magnetlabor Dresden, Forschungszentrum
Rossendorf, 01328 Dresden, Germany}
\author{A. K. Kolezhuk}
\thanks{On leave from: Institute of
Magnetism, National Academy of Sciences and Ministry of Education,
03142 Kiev, Ukraine.} \affiliation{Institut f\"{u}r Theoretische
Physik, Universit\"{a}t Hannover, 30167 Hannover, Germany}
\author{J. Krzystek}
\affiliation{National High Magnetic Field Laboratory, Tallahassee,
FL 32310}
\author{R. Feyerherm}
\affiliation{Hahn-Meitner-Institut (HMI), 14109 Berlin, Germany}

\begin{abstract}

We report the low-temperature multi-frequency ESR studies of
copper pyrimidine dinitrate, a spin-$\frac{1}{2}$
antiferromagnetic chain with alternating $g$-tensor and the
Dzyaloshinskii-Moriya interaction, allowing us to test a new
theoretical concept proposed recently by Oshikawa and Affleck
[Phys. Rev. Lett. 82, 5136 (1999)]. Their theory, based on
bosonization and the self-energy formalism,  can be applied for
precise calculation of ESR parameters of $S=\frac{1}{2}$
antiferromagnetic chains in the perturbative spinon regime.
Excellent quantitative agreement between the theoretical
predictions and experiment is obtained.

\end{abstract}
\pacs{75.40.Gb, 76.30.-v, 75.10.Jm}

\maketitle Quantum fluctuations in low-dimensional magnets give
rise to a variety of exotic strongly correlated states, making
those systems an extremely attractive ground for testing various
theoretical concepts. Isotropic $S=\frac{1}{2}$ Heisenberg
antiferromagnetic (AFM) chains with uniform nearest-neighbor
exchange interactions have a spin singlet ground state; their spin
dynamics is determined by a gapless two-particle continuum of
spin-$\frac{1}{2}$ excitations (spinons), whose parameters can be
calculated using Bethe ansatz \cite{FaddeevTakhtajan81}. In an
applied magnetic field $H$ the excitation spectrum acquires
incommensurate modes but remains gapless \cite{Mueller,Stone}. In
contrast, in $S=\frac{1}{2}$ AFM chains with an alternating
$g$-tensor and the Dzyaloshinskii-Moriya (DM) interaction, a
magnetic field induces an energy gap $E_{g} \sim H^{2/3}$
\cite{Dender}, with the effective \emph{transverse staggered}
field \cite{Dender,OshikawaAffleck97} playing a key role in the
gap formation. Fairly complete theoretical understanding of this
phenomenon has been achieved by Oshikawa and Affleck
\cite{OshikawaAffleck97,AffleckOshikawa99prb} who showed that the
gapped phase could be effectively described using the sine-Gordon
quantum field theory. A rich excitation spectrum \cite{DHN75} was
predicted to be readily apparent in the response functions
\cite{Essler+}.  Recently, with the help of inelastic neutron
scattering \cite{Kenzelmann} and electron spin resonance (ESR)
techniques \cite{Asano,Asano1,ZKKF04} the existence of soliton and
breather states in sine-Gordon spin chains has been confirmed
experimentally.

ESR is traditionally recognized as one of the most powerful and
extremely sensitive tools for probing the magnetic excitation
spectrum in exchange-coupled  spin systems. Using the temperature
or magnetic field as a tuning parameter, one can obtain valuable
information on the nature of the ground state and estimate their
important physical parameters and constants. The low-dimensional
(low-D) spin systems are of particular interest for ESR (see, for
instance, Ref.\ \cite{Katsumata,Zvyagin2} and references therein).
However, lack of a proper ESR theory often made a detailed
interpretation of experimental data, and thus an accurate
comparison with proposed models rather problematic. A new
theoretical approach for calculating ESR parameters of
$S=\frac{1}{2}$ AFM chains, which is based on bosonization and the
standard Feynman-Dyson self-energy formalism, has been recently
developed by Oshikawa and Affleck \cite{OshikawaAffleck-esr}.
Importantly, the new concept avoids the Hartree-Fock
approximation, which was previously used \cite{Mori} for
interpreting ESR in exchange-coupled spin systems, but appears to
be generally invalid in 1D magnets.

In a general case the Hamiltonian of a spin-$\frac{1}{2}$
Heisenberg AFM chain in an external field is given by
\begin{eqnarray}
\label{Ham} \mathcal{H}&=&J\sum_{i} \vec{S}_{i}\cdot
\vec{S}_{i+1}+g\mu_B H \sum_{i} S_i^z +\mathcal{H}_{\delta},
\end{eqnarray}
where the first term corresponds to the isotropic Heisenberg
interaction, the second one is the Zeeman term and the third one
represents various possible small anisotropic contributions.  In
sine-Gordon spin chains the term $\mathcal{H}_{\delta}$ in
(\ref{Ham}) is dominated by the effective staggered field $h=cH$
\cite{Dender,OshikawaAffleck97} originating from the alternating
$g$-tensor and the DM interaction,
$\mathcal{H}_{\delta}=g\mu_{B}h\sum_{i} (-1)^{i}S_{i}^{x}$. If the
temperature is high enough to destroy the soliton-breather
superstructure, but sufficiently small compared to the
characteristic energy of the exchange interaction $J$, the term
$\mathcal{H}_{\delta}$ can be regarded as a perturbation. In its
absence, i.e., in an ideal $S=\frac{1}{2}$ Heisenberg AFM chain,
the resonance would occur exactly at the frequency $\hbar\omega=g
\mu_{B}H$ and  have a zero linewidth in the low-temperature limit,
due to the conservation of the total $S^{z}$.  In the
Oshikawa-Affleck (OA) theory \cite{OshikawaAffleck-esr}, the sole
effect of the symmetry-breaking perturbation
$\mathcal{H}_{\delta}$ on the Green function is to produce the
self-energy contribution $\Sigma$ whose real and imaginary part
determine the frequency shift and the linewidth, respectively.
Other possible symmetry-breaking contributions arising, e.g., from
the exchange anisotropy, can be distinguished by the
characteristic dependence of the linewidth and the frequency shift
on the magnetic field $H$ and the temperature $T$
\cite{OshikawaAffleck-esr}.  Importantly, the OA theory allows a
$precise$ calculation of the ESR parameters and their dependence
on temperature and magnetic field. Once proven, this approach can
be used for studying peculiarities of magnetic microstructure of
$S=\frac{1}{2}$ AFM chains, and investigating anisotropies of
magnetic interactions, including those originating from the
staggered field effects.

For the first time the significant changes in ESR spectra at low
temperatures were observed in the spin-$\frac{1}{2}$
antiferromagnetic chain system with alternating $g$-tensor and the
DM interaction copper benzoate \ \cite{Date+70,Okuda} (although
the data were mistakenly explained in terms of the 3D ordering
scenario). Later on, it was found \cite{Asano} that terms
$(H/T)^2$ and $(H/T)^3$ clearly dominated in the ESR linewidth and
shift of the resonance field, respectively, being consistent with
the OA predictions. Bertaina et al.\ \cite{Bertaina+04} have
recently studied the ESR linewidth behavior in the quasi-1D
antiferromagnet $\rm BaCu_2Ge_2O_7$. It appeared that an
enhancement of 3D long-range correlations in $\rm BaCu_2Ge_2O_7$
in the vicinity of the AFM transition at $T_N=8.8$~K significantly
affected the low temperature ESR behavior, which made an accurate
quantitative comparison with the theory difficult \cite{esr-bert}.
Thus one may conclude that at a \emph{quantitative} level a
convincing evidence for the validity of the low-temperature ESR
theory for $S=\frac{1}{2}$ AFM chains with alternating $g$-tensor
and the Dzyaloshinskii-Moriya interaction is still lacking.

In this Letter we report a detailed ESR study of copper pyrimidine
dinitrate ([PM-$\rm Cu(NO_{3})_{2}(H_{2}O)_{2}$]$_{n}$, PM =
pyrimidine; hereafter Cu-PM), a spin-$\frac{1}{2}$
antiferromagnetic chain with alternating $g$-tensor and the DM
interaction, which appears to be the most perfect realization of
the quantum sine-Gordon spin chain system known to date.  The
measurements were performed in the intermediate (``perturbative
spinon'') temperature regime, $E_{g} < T < J$, at several
frequencies, allowing us to check both the temperature and field
dependencies of ESR parameters, predicted by Oshikawa and Affleck
\cite{OshikawaAffleck-esr}.  The obtained data were of a
sufficient quality to perform an accurate \emph{quantitative}
comparison with theoretical predictions, achieving an excellent
agreement between the theory and experiment. Importantly, our
results are fully consistent with the previous analysis
\cite{ZKKF04} based on the study of the frequency-field
dependencies of ESR excitations in Cu-PM in the soliton-breather
regime.

Cu-PM crystallizes in a monoclinic structure belonging to the
space group $C2/c$ with four formula units per unit cell
\cite{Feyerherm}. The lattice constants obtained from the
single-crystal X-ray diffraction are $a=12.404$~\AA,
$b=11.511$~\AA, $c=7.518$~\AA, $\beta=115.0^{\circ}$. The Cu ions
form chains  running parallel to the short $ac$ diagonal.  The Cu
coordination is a distorted octahedron, built from an almost
square N-O-N-O equatorial plane and two oxygens in the axial
positions. In this approximately tetragonal local symmetry, the
local principal axis of each octahedron is tilted from the $ac$
plane by $\pm 29.4^{\circ}$.  Since this axis almost coincides
with the principal axis of the $g$-tensor, the $g$-tensors for
neighboring Cu ions are staggered. Magnetic properties of Cu-PM
are well characterized
\cite{Feyerherm,Asano2,Wolter+03,Wolter+04,ZKKF04}. The exchange
spin interaction $J$ was found to be 36 K \cite{Feyerherm}, while
the staggered field parameter, $c$, was estimated as 0.08
\cite{ZKKF04}. Recently, the excitation spectrum of Cu-PM in the
soliton-breather regime ($T\ll E_{g}$) has been studied in detail
\cite{ZKKF04}, using the tunable-frequency ESR technique
\cite{ZvyaginKrzystek02}. Signatures of soliton and three breather
branches have been clearly identified, and the field-induced gap
has been observed directly.  In our experiments high quality
single crystals were used, with the magnetic field applied along
the $c''$ direction providing a maximum staggered field
\cite{Feyerherm,Asano2}. The ESR absorptions were fit using the
Lorentzian formula for the line shape. Usually, ESR probes the
imaginary part of dynamical susceptibility, $\chi(q,\omega)$,
responsible for the absorption. However, even a small admixture of
the real component of $\chi(q,\omega)$ (responsible for the
dispersion) can significantly distort the resonance lineshape and
thus can be an additional source of experimental error. To
minimize the contribution of the dispersive component in the ESR
spectra, special steps associated with a careful control and
correction of the microwave phase were made. An  accuracy of
better than  $5\%$ and $0.2\%$  was achieved for calculating ESR
linewidth and $g$-factor, respectively.

\begin{figure}
\begin{center}
\vspace{0cm}
\includegraphics[width=0.4\textwidth]{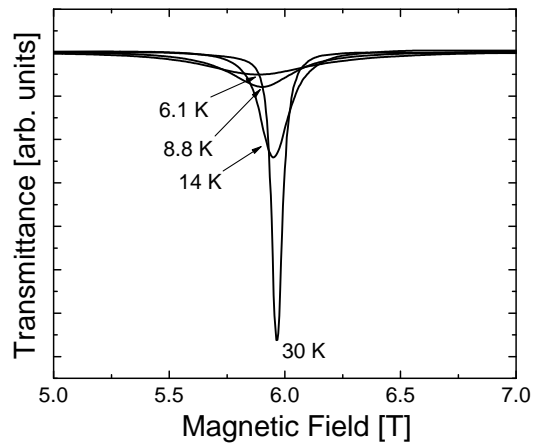}
\vspace{0cm} \caption{\label{fig:spectrum} The ESR transmission
spectra of Cu-PM  taken at a frequency of $184$~GHz at four
different temperatures. }
\end{center}
\end{figure}

\begin{figure}
\begin{center}
\vspace{0cm}
\includegraphics[width=0.45\textwidth]{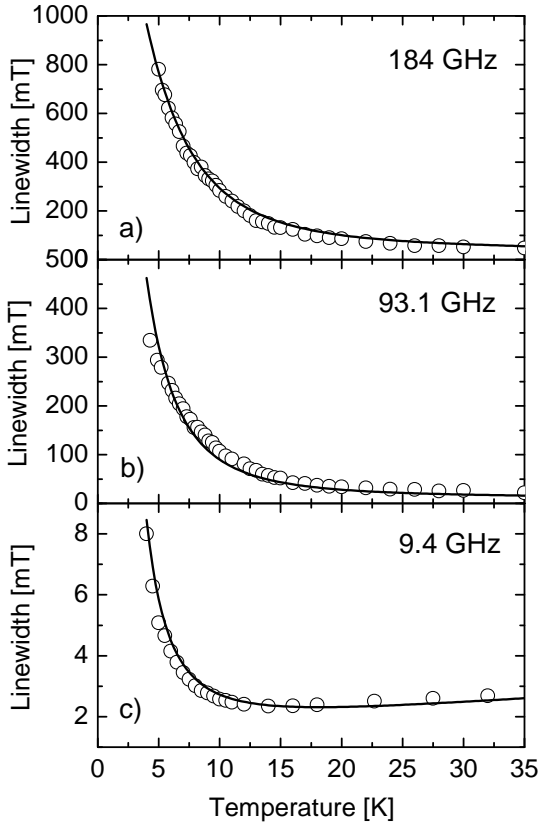}
\vspace{0cm} \caption{\label{fig:lw} The temperature dependence of
the ESR linewidth in Cu-PM at frequencies of $184$, $93.1$, and
$9.4$~GHz. Symbols denote the experimental results, and solid
lines correspond to the best global fit using Eq.\
(\protect\ref{width}).}
\end{center}
\end{figure}

Typical absorption spectra in Cu-PM at frequency of $184$~GHz at
four different temperatures are shown in Fig.\ \ref{fig:spectrum}.
One can see that upon cooling the high-temperature $\omega = gH$
resonance becomes broader, while the absorption maximum shifts
towards lower fields. Such a behavior is consistent with that
observed earlier in copper benzoate \cite{Asano} and $\rm
BaCu_2Ge_2O_7$ \cite{Bertaina+04}. In our experiments the
temperature dependence of the linewidth has been studied at three
frequencies, $9.4$, $93.1$, and $184$~GHz.  The corresponding data
are summarized in Fig.\ \ref{fig:lw}, together with results of the
OA theory.  To fit the data, the following expression
\cite{OshikawaAffleck-esr} has been used for the ESR linewidth
$\Delta H$:
\begin{equation}
\label{width} \quad \Delta H = \eta_{0} + Fz\,H\,\mbox{Im}(G) ,
\end{equation}
where $z=\Gamma(\frac{1}{4})/\Gamma(\frac{3}{4})$ and $\Gamma(x)$
denotes the gamma function, and
\begin{eqnarray*}
 G(H,T)&=&\Gamma\Big(\frac{1}{4}-i\frac{g\mu_{B}H}{2\pi T}\Big)/
\Gamma\Big(\frac{3}{4}-i\frac{g\mu_{B}H}{2\pi T}\Big),\\
 F(H,T)&=&c^{2}\sqrt{\pi/128} (J/T)\ln^{1/2}(\lambda J/T).
\end{eqnarray*}
The constant $\lambda$ in the leading log is to be viewed as a
free phenomenological parameter compensating for the effects of
subleading logarithmic terms \cite{OA-rem}. Subleading logs were
recently exactly determined for an exchange anisotropy
perturbation \cite{Maeda+05}, but for the staggered field case
their exact form is not known. The overall correction $\eta_{0}$
takes into account that the ESR linewidth is in fact \emph{finite}
in the high-temperature ($T\gtrsim J $) regime: According to the
OA theory, for $T\gg J$ the linewidth should become
temperature-independent,
\begin{equation}
\label{shift-ht}\Delta H \mapsto \eta_{0}=\alpha+\beta (g\mu_{B}H/J)^{2},
\end{equation}
finite $\alpha$ being caused by the DM interaction, and $\beta$
containing contributions both from the DM interaction and from
staggered $g$-tensor \cite{OshikawaAffleck-esr}.

\begin{figure}
\begin{center}
\vspace{2.7cm}
\includegraphics[width=0.5\textwidth]{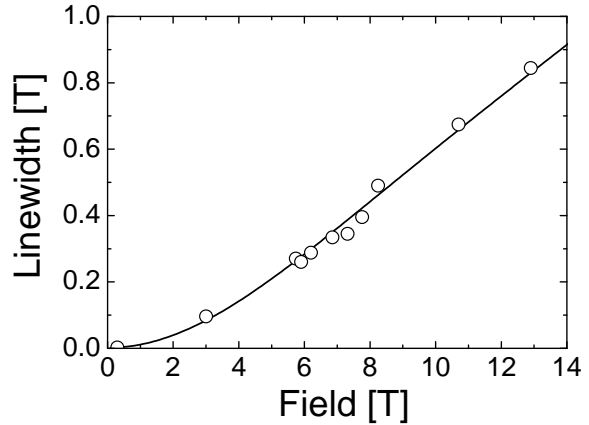}
\vspace{-3cm} \caption{\label{fig:lwH} The field dependence of the
ESR linewidth in Cu-PM at  $T=10.3$~K.
The solid line corresponds to  Eq.\ (\protect\ref{width}) with the parameters
obtained from the fit as described in the text. }
\end{center}
\end{figure}

We obtain a very good fit to the \emph{entire set} of the
linewidth data, as shown in Fig.\ \ref{fig:lw}, with the following
values of the parameters: $c=0.083\pm0.001$, $\lambda=2.0\pm0.05$,
$\alpha=(1.5\pm0.2)$~mT, and $\beta=(830\pm25)$~mT. It should be
remarked that inclusion of log corrections turns out to be
important: neglecting them makes it impossible to achieve a
\emph{uniformly} good fit for different frequencies. The obtained
staggered field parameter $c$ is in excellent agreement with the
value $c=0.08\pm 0.002$ found by us earlier from the analysis of
the frequency-field dependence of ESR modes in Cu-PM in the
soliton-breather regime \cite{ZKKF04}. The field dependence of the
linewidth measured in fields up to 13 T at temperature of 10.3 K
is shown in Fig.\ \ref{fig:lwH}, together with the results of
calculations using parameters obtained as described above. The
data are in excellent agreement with the calculated values.

As seen in Fig.\ \ref{fig:lw}(c), at $9.4$~GHz the temperature dependence of the
linewidth contains a tiny (note the scale) contribution linearly growing in
$T$. According to OA, this can be caused by a small exchange anisotropy
$\delta$, yielding the linewidth contribution $(4/\pi^{3})(\delta/J)^{2}T$ if
one assumes the anisotropy axis to be parallel to the applied field
\cite{OshikawaAffleck-esr}. Our data at $9.4$~GHz allow us to estimate this
anisotropy $(\delta/J)$ as being around $1.2\%$ which is consistent with the
susceptibility data estimating it to be within $2\%$ \cite{Feyerherm}.

\begin{figure}
\begin{center}
\vspace{2cm}
\includegraphics[width=0.5\textwidth]{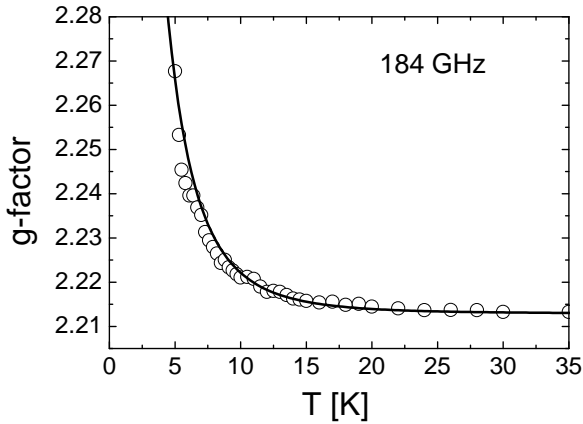}
\vspace{-3.2cm} \caption{\label{fig:g} The temperature dependence
of the effective $g$-factor in Cu-PM for a frequency of 184 GHz.
The solid line
corresponds to Eq.\ (\protect\ref{shift}) with the parameters obtained from the
linewidth fit as described in the text.}
\end{center}
\end{figure}

Let us now discuss the behavior of $g$-factor. It is worth to
mention here that while the linewidth in Cu-PM can be relatively
easily measured, an accuracy of measurements becomes really an
important problem when measuring the $g$-factor dependence,
particularly at low frequency. Employing the high-frequency and
-field ESR techniques enabled us to achieve much better resolution
than conventional low-frequency methods \cite{Asano2}, and thus to
significantly improve the accuracy of measurements. The
experimental data for the dimensionless resonance frequency shift
$\Delta g$ (the $g$-factor shift with respect to the $g$ value at
room temperature, $g=2.215$) for a frequency of $184$~GHz are
presented in Fig.\ \ref{fig:g}, together with the theoretical
curve described by the expression \cite{OshikawaAffleck-esr}
\begin{equation}
\label{shift} \Delta g = Fz\big( z- \mbox{Re}(G) \big).
\end{equation}
\emph{The same parameter values} as obtained from fitting the
linewidth data has been used.  An excellent agreement with the
theory can be seen.  For the chosen value of $H$ the breather gap
$E_{g}\approx 6$~K, which shows that the OA theory actually works
fairly well down to $T\sim E_{g}$.

In summary, we have presented a detailed study of the temperature
and field evolution of the ESR spectrum in Cu-PM, a material which
is considered as the best available realization of
spin-$\frac{1}{2}$ Heisenberg chain with alternating $g$-tensor
and the Dzyaloshinskii-Moriya interaction, in the perturbative
spinon regime. The data were analyzed in frame of the
Oshikawa-Affleck quantum-field theory \cite{OshikawaAffleck-esr}.
An excellent $quantitative$ agreement with the theory has been
achieved. The results are fully consistent with the previous
analysis \cite{ZKKF04} based on the study of the ESR excitation
spectrum of Cu-PM in the soliton-breather regime.

\emph{Acknowledgments.--} The authors express their sincere thanks
to C.L. Broholm, F.H.L. Essler,  A. Honecker, H.-H. Klauss, J.L.
Musfeldt,  M. Oshikawa, P. Schlottmann, G. Tetel'baum and  A.U.B.
Wolter for fruitful discussions. We would like to thank P.J.
Desrochers for critical reading of the manuscript and  useful
comments. The experimental part of this work was supported by NSF
Cooperative Agreement No. DMR-0084173 and by the State of Florida.
AK is supported by the Heisenberg Fellowship of Deutsche
Forschungsgemeinschaft.


\begin{thebibliography}{10}
\bibitem{esr-bert} Not underestimating the importance of the work  of Bertaina $et~al.$ \cite{Bertaina+04},
we have to mentnion here that there is a definite deviation of the
calculated values from the experimental results for  the
low-temperature part of the linewidth temperature dependencies
measured at frequencies of 34~GHz and 70.6~GHz (see Fig.\ 2b of
Ref. \cite{Bertaina+04}). A substantial inconsistency in the
extracted values of the parameter $c$ has been observed: it was
found that the OA calculations fit relatively well to the ESR data
giving $c=0.09$, which is about $35\%$ less than the value
$c=0.14$ found from the analysis of the frequency-field dependence
of magnetic excitations in the AFM long-range ordered phase.
Finally, the subtraction of the 3D contribution undertaken in
Ref.\ \cite{Bertaina+04} requires additional assumptions not
contained in the OA theory.
\bibitem{OA-rem} Note that the correct power of the leading log term is
$\frac{1}{2}$ and not $1$ as in Ref.\ \cite{OshikawaAffleck-esr}
(formulas presented in Ref.\  \cite{OshikawaAffleck-esr} contain a
misprint \cite{Oshikawa-private}).
\end{thebibliography}
\end{document}